\documentclass[submit]{EPS}
\usepackage{rotating,latexsym,amssymb}           
\usepackage{mathptmx} 
\usepackage{graphicx}
\usepackage{times}
\usepackage{natbib}
\usepackage{lineno}       
\usepackage{color}

\newcommand{\honda}{45P/H-M-P}

\newcommand{\hondaa}{45P/H-M-P }

\title{Modelling cometary meteoroid  stream traverses of the Martian Moons eXploration (MMX) spacecraft en route to Phobos}
\author{
Harald Kr\"uger, MPI f\"ur Sonnensystemforschung, G\"ottingen, Germany; PERC, Chiba Institute of Technology, Narashino, Japan, krueger@mps.mpg.de \\
Masanori Kobayashi, PERC, Chiba Institute of Technology, Narashino, Japan \\
Peter Strub,  MPI f\"ur Sonnensystemforschung, G\"ottingen, Germany; Institut f\"ur Raumfahrtsysteme, Universit\"at Stuttgart, Germany \\ 
Georg-Moragas Klostermeyer, Institut f\"ur Raumfahrtsysteme, Universit\"at Stuttgart, Germany \\  
Maximilian Sommer, Institut f\"ur Raumfahrtsysteme, Universit\"at Stuttgart, Germany \\
Hiroshi Kimura, PERC, Chiba Institute of Technology, Narashino, Japan \\
Eberhard Gr\"un,   Max-Planck-Institut f\"ur Kernphysik, Heidelberg, Germany; LASP, University of Colorado, Boulder, CO, USA \\
Ralf Srama, Institut f\"ur Raumfahrtsysteme, Universit\"at Stuttgart, Germany; Baylor University, Waco, TX, USA
}

\abstract{
The Martian Moons Exploration (MMX) spacecraft is a JAXA mission  to Mars and its moons Phobos and Deimos. MMX will 
be equipped with the Circum-Martian Dust Monitor (CMDM) which is a newly developed light-weight ($\mathrm{650\,g}$) 
large area ($\mathrm{1\,m^2}$) dust impact detector.
Cometary meteoroid streams (also referred to as trails) exist along the orbits of comets, forming fine structures of the interplanetary dust cloud.
The  streams consist predominantly of the largest cometary particles (with sizes of approximately $\mathrm{100\,\mu m}$ to 1~cm) which are 
ejected at low speeds and 
remain very close to the comet orbit for several revolutions around the Sun. 
The Interplanetary Meteoroid Environment for eXploration (IMEX) dust streams in space model is a new and recently published universal model for cometary meteoroid streams in the inner Solar System. We use IMEX to study the detection conditions of cometary dust  stream particles with CMDM during the MMX
mission in the time period 2024 to 2028.
The model predicts traverses of 12 cometary meteoroid  streams with  
fluxes of $\mathrm{100\,\mu m}$  and 
bigger particles of at least $\mathrm{10^{-3}\,m^{-2}\,day^{-1}}$ during a total time period of approximately 90~days.  
The highest flux of $\mathrm{0.15\,m^{-2}\,day^{-1}}$
is predicted for comet 114P/Wiseman-Skiff in October 2026.
With its large detection area and high sensitivity CMDM will  be able to detect  cometary meteoroid 
 streams en route to Phobos. Our simulation results for the Mars orbital phase of MMX also predict the
occurrence of meteor showers in the Martian atmosphere which may be observable from the Martian surface
with cameras on board landers or rovers. Finally, the IMEX model can be used to study the impact hazards  
imposed by meteoroid impacts onto 
large-area spacecraft structures that will be particularly necessary for  crewed deep space missions. 
}


\begin{document}

\maketitle

\bibliographystyle{apalike}

\section{Introduction}

The Japanese Space Exploration Agency (JAXA) is planning to send the Martian Moons Exploration (MMX)
spacecraft to the Martian moons Phobos and Deimos and  to bring samples from Phobos to Earth 
\citep{kuramoto2018}. The launch of MMX is planned for 2024, 
and Phobos sample arrival at Earth is scheduled for 2029.  

MMX will be equipped with the large area Circum-Martian Dust Monitor (CMDM) which has a sensitive area of 
$\mathrm{1\,m^2}$ \citep{kobayashi2018a}. The primary goals of the  CMDM  dust investigations are
to understand dust processes in the circum-Martian environment, including  the search for a proposed but still undiscovered 
circum-Martian dust ring 
\citep{soter1971,hamilton1996b,krivov1997,ishimoto1997,showalter2006}. A second goal is the measurement of  
interplanetary and interstellar dust \citep{kobayashi2018a} as well as cometary meteoroid streams. 

The prominent dust tail of a comet predominantly consists of small submicrometer-sized  particles that are blown out by the solar radiation 
pressure force. Larger dust particles with sizes of approximately 
$\mathrm{100\,\mu m}$ to 1~cm  are ejected from the cometary nucleus at low speeds and remain very close to the comet 
orbit for several revolutions around the Sun \citep{agarwal2010}. They slowly spread in the comet orbit as a 
result of small differences in orbital period, and they form a tubular structure along the  orbit of the parent comet filled with dust, 
called a dust  stream or trail. Particles in the intermediate size range from approximately $\mathrm{1\,\mu m}$ to $\mathrm{100\,\mu m}$ are still 
subject to radiation pressure, thus migrating away from the comet's orbital position, but they stay
on trajectories similar to the orbit of their parent comet for some time.

Dust trails in the vicinity of comets were first observed by the Infrared Astronomical 
Satellite \citep[IRAS;][]{sykes1986,sykes1992}. In 
subsequent infrared observations, 
at least 80\% of the observed Jupiter-family comets were associated with dust trails  which can thus be 
considered one of their generic features \citep{reach2007}. 
A recent review of the present knowledge on cometary dust, including dust trails, was  given by \citet{levasseur-regourd2018}.

These meteoroid trails form fine structures superimposed on the interplanetary 
background dust cloud. 
When the Earth intercepts a cometary trail, the
 particles collide with the atmosphere and show up as meteors and fireballs 
  with hourly rates significantly enhanced over the  sporadic meteor background flux
 \citep[][and references therein]{koschny2019}. 
 Effects of meteoroid impacts were also observed on the Earth's Moon and on other planets \citep[see][for a comprehensive review]{christou2019}. 
  Meteor trails can impose safety risks for Earth orbiting and interplanetary spacecraft. Although the impact 
 probability per spacecraft is small, spacecraft anomalies  very likely related to meteoroid impacts were observed in the past
 \citep{caswell1995,cameron2004}. The risk is considered sufficiently high to occasionally affect crewed spaceflight operations
  \citep{beech1995}.

 In the past, 
 the detection of meteoroid trails by spacecraft-based in situ dust detectors was hindered by the relatively small sensitive
 area of the instruments and low particle fluxes in the trails. Only recently could a small number of impacts of likely cometary 
 meteoroid particles  be identified in the data set collected by the in situ dust instruments onboard the Helios spacecraft 
 in the inner Solar System in the 1970s \citep{altobelli2006,krueger2020}. 
Interestingly, the identification of trail particles was 
only possible because Helios  traversed the same region of space and thus, the same cometary trails, many times which 
significantly increased the chances to detect trail particles with these relatively small impact ionization 
dust sensors. 
MMX with its large area  CMDM dust detector will be excellently equipped to make the first undisputable in situ detections of  
cometary meteoroid trails. 

Cometary dust trails can be simulated with the Interplanetary Meteoroid Environment for eXploration (IMEX) dust streams in 
space model which was  developed under contract by the European Space Agency  \citep{soja2015a,soja2015b,soja2019}. 
IMEX is a new universal and physical model for dust dynamics and orbital evolution  that simulates 
cometary dust trails in the 
inner Solar System. The  model follows the trails 
of 420 comets and is ideal for studying meteor streams and cometary dust trails as measured by in situ detectors 
and observed in infrared images. A recent review of presently existing up to date  models for dust in 
interplanetary space in the inner 
solar system can be found in \citet{gruen2019}.

We use the IMEX model to study future cometary trail traverses by the MMX spacecraft during its 
interplanetary transfer to Mars and during its Martian orbital phase. After a 
brief description of  the MMX mission and the CMDM instrument, we give an overview of the IMEX model. Next, we present 
and discuss the results of our IMEX simulations and, 
finally, we summarise our conclusions.

\section{MMX Mission and Circum-Martian Dust Monitor}

\label{sec:mission}

Building on the experience of the asteroid sample return spacecraft Hayabusa and Haya\-busa~2, JAXA is planning to 
launch the Martian Moons eXploration (MMX) spacecraft to Phobos and Deimos \citep{kuramoto2018,kuramoto2019}. The 
mission will perform the 
first landing on one of the Martian moons, Phobos, collect samples from its surface, and return them to 
Earth. In addition to the returned samples, the mission will 
perform close-up remote sensing and in situ measurements at the two  moons, and deploy a small rover on 
the Phobos surface. The launch of MMX is planned for 2024, the spacecraft will be in Mars orbit from 2025 to 2028, 
and the return module will arrive at Earth in 2029. 

For its  transfer to Mars  MMX will use a low-energy transfer orbit, which takes about
one year to reach the planet. This trajectory takes more time than a classical Hohmann transfer but significantly reduces the 
amount of fuel needed \citep{topputo2015,ogawa2019}. After  Mars orbit insertion, MMX will initially orbit the planet on 
a trajectory near Phobos' orbit.
Subsequently the spacecraft will be injected into a quasi-satellite orbit around the moon with a minimum altitude above the Phobos surface 
of 20~km or less \citep{kuramoto2019}. 
MMX will be a 3-axis stabilised  spacecraft throughout the entire mission.

One of the scientific instruments on board MMX will be the Circum-Martian Dust Monitor \citep[CMDM;][]{kobayashi2018a}.
It is a large area dust detector with $\mathrm{1\,m^2}$ sensor area. For dust detection
 CMDM will use the outermost polyimide film of the spacecraft multi-layer insulation (MLI), 
which is a thermal blanket of the spacecraft.  The sensor will use the MLI on one flat side of the spacecraft and have 
a field-of-view of $2\pi$. Additional piezoelectric  elements will be mounted on the MLI surface. A dust  
particle hitting or penetrating the MLI surface generates a stress wave that subsequently propagates through the MLI film. It can be 
detected by one or more of the piezoelectric sensors, and the position of the impact on the MLI  can be derived 
from the differences between the arrival times of the signals measured on the multiple pick-up sensors. From the derived position 
of the crater or penetration location, it will be possible to distinguish whether it is a dust impact or  noise event \citep{kobayashi2016}. 
Information about the projectile's mass and impact speed 
can be derived from the waveform of the signal.
At $\mathrm{10\,km\,s^{-1}}$ 
 impact speed the instrument can detect impacts of particles as small as approximately 
$\mathrm{3\, \mu m}$ in radius. 

\section{IMEX cometary trails model}

\label{sec:model}

In order to identify time intervals when the MMX spacecraft will traverse cometary meteoroid trails, we use
the IMEX dust streams in 
space model 
\citep{soja2015a,soja2015b,soja2019}.  The model generates  trails for 362 Jupiter-family, 
40 Halley-type, and 18 Encke-type comets available in the JPL 
Small Body Database (SBDB) as of 1 August 2013, which have perihelion distances $q < 3$ AU, semimajor axes $a <$ 30 AU, and 
defined total visual magnitudes. 

Particles are emitted when the comet is in the inner Solar System, taking into account 
 perihelion passages between the years 1700 and 2080  for Encke-type comets, and between 1850 and 2080 for 
Jupiter-family and Halley-type comets, respectively. This reflects the fact that the most recently released dust is expected 
to be most important, and also the maximum size of the database that could be maintained at the time when the model was developed. 

For each passage through the inner Solar System  within 3~AU of the Sun of each comet (which we refer to as apparition in the following),
particles are emitted from the sunlit hemisphere of the comet nucleus within the time ranges specified above. 
About 28,000 particles are ejected per comet per apparition for Halley-type comets; and  about 14,000 
for other comets.

The dust ejection is described by the comet emission model of 
 \citet{crifo1997}. The model assumes the dust emission to be driven by  water gas production within 3~AU
 of the Sun. It estimates the water production rate  
using the visual magnitude, and a gas-to-dust ratio \citep{jorda2008}.
The JPL Small Body Database provides total and nuclear magnitudes.

Dust-to-gas mass ratios can be estimated for individual comets, and they mostly range from 0.1 to 3, 
though higher values are possible. They appear to be dependent on heliocentric distance \citep{ahearn1995}. 
Given the large uncertainties in dust-to-gas ratios, the model uses a value of 1. 
Deviations from this can be considered in the analysis of individual comets.

The IMEX model uses the mass distribution model of \citet{divine1987} and \citet{agarwal2007a,agarwal2010}, with model parameters
given by \citet{soja2015a}. 
The  mass distribution covers the  range from 
$\mathrm{10^{-8}\,kg}$ to $\mathrm{10^{-2}\,kg}$, separated into eight mass bins \citep[approximately corresponding to 
$\mathrm{100\, \mu m}$ to 1~cm particle radius;][]{soja2015a}. 
The particle density is assumed to be $\mathrm{\rho = 1,000\,kg\,m^{-3}}$. For comets with  unknown radius a value of 
1~km is assumed \citep{soja2015a}.

The trajectory of each emitted particle is integrated individually including solar gravity, 
planetary perturbations, solar radiation pressure and Poynting-Robertson drag. 
The
model calculates the impact velocity for each individual particle onto the 
spacecraft as well as dust number density and flux. A detailed model description including an 
application to the  trail of comet 67P/Churyumov-Gerasimenko was given by \citet{soja2015a}. 
We use the IMEX model to identify time intervals when MMX will traverse the 
meteoroid trails of comets between September 2024 and August 2028 which covers the 
presently planned interplanetary transfer to Mars and the Martian 
orbital phase of the spacecraft.

For the interplanetary transfer from Earth to Mars we use a preliminary MMX trajectory provided by ISAS/JAXA. It is the
nominal trajectory in the mission analysis performed by \citet{campagnola2018} 
with a launch in 2024. For the Martian orbital phase of MMX we use Mars as the target for our simulations. 
We do not consider the interplanetary return
 to Earth  in this work.  CMDM will be mounted on the spacecraft module that will return to Earth and be able to 
 collect data during this mission phase, however, we do not
 have a trajectory for this mission phase available at the present time. Given that IMEX is a time-dependent model, the use
 of the actual spacecraft trajectory is a prerequisite to make reliable predictions for the occurrence of trail traverses and 
 the dust fluxes during  such
 traverses. 

\section{Results}

\label{sec:results}

Figure~1 shows the simulated dust fluxes  during  cometary trail traverses of MMX along its interplanetary
transfer  to Mars. The MMX trajectory for this mission phase is shown in Figure~2, and  details of these trail traverses 
are given in Table~\ref{tab:results}. 

The highest flux during the transfer to Mars is predicted for comet 21P/Giacobini-Zinner: For a trail traverse in October 2024 
the model predicts a maximum flux of  
$\mathrm{8\cdot 10^{-3}\,m^{-2}\,day^{-1}}$. 
It is the highest meteoroid trail flux expected during the interplanetary transfer phase of MMX. 
For comets 103P/Hartley, 24P/Schaumasse 
and   45P/Honda-Mrkos-Pajdu\v{s}{\'a}kov{\'a} (hereafter \honda) the predicted fluxes are approximately  
 $\mathrm{10^{-3}\,m^{-2}\,day^{-1}}$ in late October 2024, mid December 2024 and in May 2025.  
For these four comets the time intervals  with fluxes of approximately  $\mathrm{10^{-3}\,m^{-2}\,day^{-1}}$ or higher last between  two 
(24P/Schaumasse and 103P/Hartley) and 12 days (\honda), adding up to a total of 20~days. 
  The model finds 23 additional comets with   fluxes below  $\mathrm{10^{-3}\,m^{-2}\,day^{-1}}$
which  we do not consider relevant. MMX will cross the trail of  \hondaa twice, namely in  September 2024 and in  May 2025. 

Figure~2 shows the locations of trail traverses and the impact directions of trail particles 
 in the spacecraft-centered reference frame projected onto the ecliptic plane. 
The particle impact speeds for the above four comets are in the range
 $\mathrm{10\,km\,s^{-1}} \lesssim |v_{\mathrm{imp}}| \lesssim \mathrm{20\,km\,s^{-1}}$. Table~\ref{tab:results} gives the 
 impact speeds 
 together with the impact directions of the particles. The latter are given as the upstream direction  in the spacecraft 
 frame of reference of the particle flow which
corresponds to the radiant of a meteor stream in a planetary atmosphere.

For the MMX Mars Orbital Phase the 
simulated dust fluxes are shown in Figure~3.
For seven comets with a total of nine trail  traverses the predicted fluxes exceed 
 $\mathrm{10^{-3}\,m^{-2}\,day^{-1}}$, see also Table~\ref{tab:results}. The time intervals with
 fluxes above this limit last between  2 and 15 days, adding up to a total of 70~days. The comet trail traverse with the highest predicted flux is the one of 114P/Wiseman-Skiff
 with a maximum  of  $\mathrm{1.5\cdot 10^{-1}\,m^{-2}\,day^{-1}}$. However, this trail is rather collimated, leading to a short traverse which lasts  only 
three days. 
  The model finds 72 additional comets with   fluxes below  $\mathrm{10^{-3}\,m^{-2}\,day^{-1}}$. 
As for the interplanetary transfer phase, we do not consider them relevant. 

Figures~1 and 3 also
show the flux of sporadic interplanetary particles predicted by the Interplanetary Meteoroid Engineering Model (IMEM) developed by 
\citet{dikarev2005a}. For the IMEM simulations we assumed a spinning flat plate sensor. Given that the actual CMDM orientation on board the 
3-axis-stabilized MMX spacecraft during the mission is not known at the present time, this is a reasonable approximation. 
We adopted the same particle size 
range as for the  IMEX meteoroid trail simulations
($\mathrm{10^{-8}\,kg}$ to $\mathrm{10^{-2}\,kg}$) so that the fluxes can be directly compared with the fluxes predicted for the meteoroid
trail traverses (Due to artefacts in the  predicted fluxes for a lower mass cutoff at $\mathrm{10^{-8}\,kg}$ caused by
the parametrisation of the cometary particle populations in the IMEM model, we used a slightly lower boundary of $\mathrm{4\cdot 10^{-9}\,kg}$ 
for the simulations 
and scaled the curves shown in Figures~1 and 3  by a factor of 2.5 that we derived from the particle size distribution). The highest predicted trail 
fluxes typically exceed the flux of sporadic interplanetary particles of the same size by approximately
one (during interplanetary transfer) to two orders of magnitude (during Mars orbital phase). We ignore interstellar dust because it does not make any
significant contribution to the dust fluxes in this size range \citep{strub2019}. Fluxes expected  for smaller interplanetary and interstellar
particles down to the detection limit of CMDM as predicted by IMEM are given by \citet{kobayashi2018a}. 

As for the interplanetary transfer,  
the locations of trail traverses and  impact directions of trail particles  
are shown in Figure~4. The particle impact speeds are in the range
 $\mathrm{10\,km\,s^{-1}} \lesssim |v_{\mathrm{imp}}|  \lesssim \mathrm{22\,km\,s^{-1}}$. Again the impact directions 
 and speeds are given in Table~\ref{tab:results}. Strictly speaking, impact speeds and directions refer to an observer on Mars because 
 we have used Mars as the target for our simulations. The orbital elements of the identified comets are 
given in Table~\ref{tab:comets}  in the Appendix. All comets identified in our analysis are Jupiter Family Comets with orbital 
periods between 5.2 and 13.8~years. 

Column~6 in Table~\ref{tab:results} lists the total number of expected particle detections with CMDM during the
traverse of each comet's trail (fluence). All these numbers are below 1, with the highest values being the fluences of 
approximately 0.3 for comets 19P/Borrelly and 114P/Wiseman-Skiff. It has to be noted that these fluences refer to particle sizes of 
approximately $\mathrm{100\, \mu m}$ and bigger, while CMDM can detect particles at least one order of magnitude smaller. 
These smaller particles are not covered by the IMEX model and, thus, the  fluences detectable with CMDM are expected to
 be significantly higher. We will discuss this in the next Section.

\begin{sidewaystable}[tbp]
\small
      \caption{Results from the IMEX cometary dust trails simulations for MMX. The time interval of the trail 
      traverse together with the predicted maximum flux (of  $\mathrm{100\,\mu m}$ and bigger particles) are given in columns (2) to (5).  Column (6)
      gives the fluence on to a $\mathrm{1\,m^2}$ detector detectable during the time interval given in column (2).
      Columns (7) to (10) give the speed vector, while columns (11) and (12) give the impact direction of the particles in ecliptic coordinates 
(opposite to speed vector), 
both  in the spacecraft-centered reference frame. The impact direction corresponds to the radiant of a meteor stream in a planetary atmosphere.}
   \begin{tabular}{@{} lccccccccccc @{}} 
   \hline
\multicolumn{1}{c}{Comet} & Time interval with flux              & Duration & Day with    &Maximum &  Fluence  & \multicolumn{4}{c}{Impact speed at maximum flux} &\multicolumn{2}{c}{Impact direction} \\
                              & $\mathrm{\geq 10^{-3}\,m^{-2}\,day^{-1}}$ &           &maximum flux &    flux&                &   $|v_{\mathrm{imp}}|$  &  $v_x$ &  $v_y$ &$v_z$ & $\mathrm{\lambda_{ecl}}$& $\mathrm{\beta_{ecl}}$\\
                          &                                         & [days] &         & [$\mathrm{m^{-2}\,day^{-1}}$]& [$\mathrm{m^{-2}}$]& [$\mathrm{km\,s^{-1}}$]&[$\mathrm{km\,s^{-1}}$]&[$\mathrm{km\,s^{-1}}$]&[$\mathrm{km\,s^{-1}}$]&[$^{\circ}$] &[$^{\circ}$]\\
\multicolumn{1}{c}{(1)}  &      (2)          &           (3)       &    (4)     &      (5)                   &    (6) &   (7)  &   (8)  &     (9)                &    (10) & (11) & (12)\\
       \hline
       \multicolumn{12}{c}{Interplanetary Transfer Phase} \\
21P/Giacobini-Zinner & 04.10.2024 - 11.10.2024    & 8   & 05.10.2024        & 0.0082 & 0.013 &    20.2                    &  1.4   &  0.9   & -20.1  &     212                &     85   \\ 
24P/Schaumasse       & 14.12.2024 - 16.12.2024    & 2   & 16.12.2024        & 0.0011 & 0.002 &    20.4                    & -5.6   & -17.1  &   9.6  &      72                &    -28   \\
\honda               & 05.05.2025 - 17.05.2025    & 13  & 17.05.2025        & 0.0013 & 0.015 &    17.6                    &-17.4   &  2.7   &  0.0   &     351                &      0   \\
103P/Hartley         & 25.10.2024 - 26.10.2024    & 2   & 26.10.2024        & 0.0009 & 0.001 &    10.0                    & -7.0   &  1.1   &  -7.1  &     351                &     45   \\
        \multicolumn{12}{c}{Mars Orbital Phase} \\
3D/Biela                 & 30.05.2026 - 15.06.2026      & 17 & 10.06.2026        & 0.0064 & 0.006 &    20.5                    & -18.5  & -8.0   & -3.8   &     23                &    11   \\ 
3D/Biela                 & 13.06.2027 - 25.06.2027      & 13 & 20.06.2027        & 0.0045 & 0.050 &    20.4                    & -18.5  & -7.8   &  3.6   &     23                &    -10   \\
19P/Borrelly             & 08.09.2026 - 18.09.2026      & 11 & 15.09.2026        & 0.0310 & 0.332 &    15.7                    & -2.4   & -0.3   & 15.5   &      8                &    -81   \\
24P/Schaumasse           & 02.09.2026 - 13.09.2026      & 12 & 09.09.2026        & 0.0039 & 0.041 &    16.4                    & -7.6   & -13.9  &  4.1   &      61                &  -14   \\
\honda                   & 21.12.2027 - 26.12.2027      & 6  & 24.12.2027        & 0.0355 & 0.120 &    22.3                    &-16.6   & 15.0   & -0.2   &     318                &    1   \\
114P/Wiseman-Skiff       & 04.10.2026 - 07.10.2026      & 3  & 06.10.2026        & 0.1535 & 0.300 &    11.0                    &-4.9    & -1.1   &  -9.9  &      13                &    63   \\
185P/Petriew             & 24.06.2026 - 05.07.2026      & 12 & 04.07.2026        & 0.0046 & 0.037 &    10.7                    & -2.3   &  3.1   & -10.0  &     306                &    69   \\
275P/Hermann             & 22.03.2027 - 25.03.2027      & 4  & 24.03.2027        & 0.0110 & 0.034 &    12.3                    & -2.6   & -5.7   & -10.6  &      66                &   59   \\
\hline
   \end{tabular}
   \label{tab:results}
\end{sidewaystable}

\section{Discussion}

\label{sec:discussion}

Individual  trail particles originating from comets \hondaa and 72P/Denning-Fujikawa were likely 
detected by the Helios dust instruments already in the 1970s 
 \citep{altobelli2006,krueger2020}.  IMEX predicts a flux of approximately $\mathrm{10^{-2}\,m^{-2}\,day^{-1}}$ for the traverses 
 of these meteoroid trails by Helios. It has to be emphasised that 
 the dust sensors on board Helios had a combined effective sensor area for trail particle detections of only about $\mathrm{30\,cm^2}$. 
 However, Helios traversed the same region of space, and thus the same comet trail, ten times, which increased the detection
 probability also by a factor of ten. Considering that the sensor area of CMDM will be larger by a factor of more than 300 while the dust
 fluxes predicted by IMEX are similar within an order of magnitude, we conclude that CMDM has a very good chance to detect cometary 
 trail particles as well.

Our analysis of the Helios dust data  showed an offset by a few days for 
some dust particle detections as compared to the times predicted by the IMEX model \citep{krueger2020}. Various reasons 
may be responsible for such an offset which  include 
shortcomings of the IMEX model like, for example, the dust ejection model employed which may not fully cover all relevant 
aspects of dust ejection, or the particle trajectory computation which follows the particle motion only for up to 300 years. 
Furthermore, the model simulates 
particles with masses $\mathrm{10^{-8}\,kg}$ and
bigger, corresponding to particle radii above approximately $\mathrm{100\, \mu m}$ while    
the masses of the particles detected by Helios as derived from the calibration of the Helios dust instruments 
were  at least four orders of magnitude  less massive, 
corresponding to particle radii of a few micrometers to about $\mathrm{10\, \mu m}$  (Note that the Helios dust 
instruments had a detection threshold for dust particles with masses of approximately $\mathrm{3 \cdot 10^{-16}\, kg}$ 
at an impact speed of $\mathrm{10\,km\,s^{-1}}$ \citep{gruen1980a}, corresponding to a particle radius of approximately $\mathrm{0.3\,\mu m}$).
Such smaller particles are more susceptible to radiation pressure and Poynting-Robertson drag 
than the larger ones which leads to an increased particle dispersion in space. From our analysis of the 
Helios data we concluded that the size 
(mass) of the impactors as derived from the instrument calibration may be significantly underestimated, even more so if cometary particles are fluffy 
aggregates as shown, for example, by the
results from the Rosetta mission to comet 67P/Churymov-Gerasimenko \citep{guettler2019,kimura2020a,kimura2020b}. 
A more detailed discussion is given by \citet{krueger2020}.
To conclude, the time intervals for the
trail traverses  given in  Table~\ref{tab:results} may be too conservative when taking into account particles smaller than
$\mathrm{100\,\mu m}$.

  
 The IMEX model simulates relatively large particle sizes $r_d \ge 100\,\mu \mathrm{m}$, much larger than the particles 
 detectable by the CMDM dust detector ($r_d \ge 3\,\mu \mathrm{m}$). However, smaller particles are known to be present in the ejecta cloud of a comet, following a differential power law index of approximately $-4$ in their size distribution \citep[][ and references therein]{agarwal2010}, which corresponds to an index of approximately $ -3$ for a cumulative size distribution. Their dynamical behaviour is not the same as for particles with $r_d\gtrsim 100\mu \mathrm{m}$ due to the increasing effect of non-gravitational forces on the smaller particles. However, for sizes $10\mu \mathrm{m} \le r_d \le 100\mu \mathrm{m}$ their orbital characteristics are sufficiently similar to the larger particles, but spatially offset due to perturbations, as preliminary tests have demonstrated. An updated version of the IMEX Streams model for smaller particle sizes is planned, but currently not available for predictions. 

In order to get to a more realistic representation of the expected fluxes, we consider two cases with a lower cutoff of $r_\mathrm{cutoff,1} = 10\,\mu \mathrm{m}$ and $r_\mathrm{cutoff,2} = 30\,\mu \mathrm{m}$, resprectively. We extrapolate from the IMEX model flux of particles with sizes $r_d \ge 100\,\mu \mathrm{m}$ using a cumulative power law size distribution index of $-3$. This leads to an increase in total flux by a factor of 31.6 in the case of a cutoff 
at $30\,\mu \mathrm{m}$, and a factor of 1\,000 for the $10\,\mu \mathrm{m}$\, cutoff. While it is unlikely that all particles in this size range remain in the stream, we consider a lower cutoff of 10 to $30\,\mu \mathrm{m}$ conservative, and therefore a reasonable estimate for the expected fluxes and fluences overall. Based on the fluences listed in Table~\ref{tab:results} (column~6), on the order of approximately  10 to 100 
trail particle impacts are expected during traverses of the densest trails  predicted by the IMEX model. This is also supported by 
our Helios observations mentioned above. 

In addition to this, a comparison of the IMEX  model to observations of cometary meteoroid streams by \citet{soja2015a}  concluded that the model likely underestimates the true fluxes by up to an order of magnitude, further increasing the odds for a successful detection of cometary streams.

 Taking these considerations into account, our results indicate that, given its large sensitive area  of $\mathrm{1\,m^2}$, CMDM has  good chances to detect
cometary meteoroid trail particles. Comets 21P/Giacobini-Zinner  and \hondaa are the best candidates for  trail detections during MMX' 
interplanetary transfer to Mars,  while comets 114P/Wiseman-Skiff, 19P/Borrelly and \hondaa are particularly 
good candidates during the spacecraft's Martian orbital phase.   
Successful detections of individual cometary trails with CMDM and  dust instruments on board other spacecraft 
will likely lead to an improved calibration of the IMEX Streams model and to more reliable flux predictions in the future.

The primary goal of the CMDM dust investigations at Mars will be the search for circum-Martian dust rings formed by particles
originating from Phobos and Deimos \citep[see, e.g.][for recent reviews]{zakharov2014,spahn2019}. Given that MMX will be orbiting 
Mars in an orbit similar to Phobos during a significant 
fraction of the mission time, the spacecraft
will largely be immersed in these rings  and the question arises how CMDM can effectively separate cometary trail
particles from ring particles. Dust number densities predicted for the Martian  rings are approximately $\mathrm{6\cdot10^{-6}\,m^{-3}}$ 
for the Phobos ring for particles larger than $\mathrm{30\,\mu m}$ 
and approximately $\mathrm{10^{-4}\,m^{-3}}$ for the Deimos ring for particles larger than $\mathrm{15\,\mu m}$ \citep{krivov1997,krivov2006}. 
Smaller particles are not expected to exist in abundance in the rings because their orbits are unstable. The expected 
particle impact speeds are in the range 0.2 to $\mathrm{0.8\,km\,s^{-1}}$ \citep{kobayashi2018a}. Assuming a typical value 
 of $\mathrm{0.5\,km\,s^{-1}}$, these number densities convert to dust fluxes of approximately $\mathrm{300\,m^{-2}\,day^{-1}}$ 
and $\mathrm{4000\,m^{-2}\,day^{-1}}$, respectively. Number densities for $\mathrm{100\,\mu m}$ and bigger particles as 
simulated by IMEX are not given by \citet{krivov1997}. However, from their results we estimate that the fluxes  of such particles in the rings 
are likely more than one (in the case of Phobos) 
up to two (Deimos) orders of magnitude lower.  Even these reduced numbers exceed the fluxes for cometary trail traverses
predicted by our IMEX simulations by a few orders of magnitude.  This makes particle identifications solely by  enhanced 
dust fluxes 
 very unlikely and, hence, we will have to use other criteria to identify cometary trail traverses. 
The impact speeds of trail particles exceed $\mathrm{10\,km\,s^{-1}}$ (Table~\ref{tab:results})
and are thus more than one order of magnitude higher than those of the ring particles. Furthermore, the impact direction of the
particles can be derived from the spacecraft orientation at the particle impact time. 
Thus, the measured impact speed together with the impact direction will be the crucial parameters to  separate cometary trail particles
 from Martian ring particles in the CMDM data set. 

Similarly,  during the interplanetary cruise phase of MMX, trail particles will have to be separated from the 
 background flux of sporadic interplanetary particle impacts. Given that the impact speed
will not serve as a crucial parameter because  both populations will have similar  speeds, we will instead
search for cometary trail particles in the data set by using statistical techniques. We will use a method 
employed by \citet{oberst1991} to identify temporal clustering of meteoroid impacts detected by the Apollo lunar seismic network.
The application of this method to in-situ dust detector data was described  by 
 \citet{gruen2001b} and successfully applied  for the identification of Jovian dust stream particles
  in the Ulysses dust detector data set \citep{baguhl1993a,krueger2006c}.

About 15 years ago there was the claim that 
a meteor originating from comet 114P/Wiseman-Skiff was photographed in the Martian sky by the Mars exploration rover Spirit 
 \citep{bell2004}. However, later work quantified the effects of cosmic ray hits on the Spirit 
Pancam as part of a dedicated meteor search and placed this detection in doubt \citep{domokos2007}. We also 
simulated the trail traverse of 114P/Wiseman-Skiff in 2004 with IMEX and we did not find a significantly enhanced 
dust flux in that period, consistent with the interpretation that the feature seen in the image taken by the 
Mars rover was a likely false detection. 

In 2014 Mars experienced a very close 
encounter with comet C/2013 A1 (Siding Spring) at a distance of approximately 135,000~km \citep{withers2014}. On this occasion 
the planet traversed  the  cometary 
coma and meteoroid stream, leading to   
perturbations in Mars' atmosphere and ionosphere. The event created a temporary planet-wide ionospheric layer below 
Mars' main dayside ionosphere:
The Neutral Gas and Ion Mass Spectrometer on board the Mars Atmosphere and Volatile EvolutioN spacecraft (MAVEN)
detected a large variety of metal ions following the ablation of dust particles from the cometary coma 
\citep[Na, Mg, Al, K, Ti, Cr, Mn, Fe, Co, Ni, Cu,  Zn, and possibly Si and Ca;][]{benna2015}. 
Intense ultraviolet emission due to magnesium and iron was attributed to dust trail particles approximately 1 to $\mathrm{100\,\mu m}$ in 
size  \citep{schneider2015}.  These observations showed that the ablation of meteoroids can lead to significant temporal 
changes in the composition of planetary atmospheres and ionospheres.

Comets with close encounters to Mars were  studied by various authors 
 \citep{adolfsson1996} \citep{treiman2000,selsis2004,christou2010a}  \citep{christou2011}, see  also \citet{christou2019}
for a recent review. These analyses studied the potential occurrences of meteor streams in the Martian atmosphere and derived 
impact speeds and radiant directions of 
the streams based on the encounter distance of the comet, but they were not able to make  predictions for dust fluxes.  In these works  comets 24P/Schaumasse, 
 \honda, 114P/Wiseman-Skiff and 275P/Hermann -- among many other comets -- were identified as potential candidates for sources 
 of meteor streams in the Martian
atmosphere. In our analysis, fluxes exceeding $\mathrm{10^{-3}\,m^{-2}\,day^{-1}}$ are predicted for these comets. 
\citet{christou2019} list eight potential parent comets 
for Martian meteor showers in the time interval 2019 to 2021, with only 3D/Biela being in our list of predicted high flux candidates in the time 
interval 2024 to 2028 considered in our work. 
 We also find many more 
comets with  lower predicted fluxes in our simulations which we do not consider further as it is beyond the scope of this paper to
make a detailed comparison of cometary trail traverses predicted by different methods. Finally, 
of the comets with high fluxes identified in our work,  at least \honda, 21P/Giacobini-Zinner and 
3D/Biela are sources of known meteor showers in the Earth atmosphere (the $\alpha$ Capricornids, the Giacobinids and the Andromedids). 
Up to now the Earth remains the only planet where meteors have been observed  
\citep{christou2019}. 

Comet 3D/Biela is a lost comet as indicated by the letter "D". It most likely disintegrated in the early 
1840s, and  two components were subsequently observed in 1846. The last observation of the two fragments occurred 
in 1852, and the comet has been lost since then \citep{kronk2003}. Comet Biela may have been the source of a very 
brilliant meteor shower in 1872, the Andromedids or Bielids.  
Nevertheless, meteoroids from this
comet are likely still forming a meteoroid trail along the comet's former orbit as implied by the fact 
that its meteor shower is still active in the present time \citep[e.g.][]{green2013}.

Comet 103P/Hartley (or Hartley 2) was the target of the Deep Impact extended (EPOXI) mission. Close-up
images of this comet's nucleus revealed centimeter to decimeter sized objects in its  vicinity \citep{ahearn2011}, 
at least a fraction of which is likely to become part of the comet's dust trail.  
Finally, in 2001 comet 19P/Borrelly was the target of the Deep Space 1 spacecraft.

Comet trails were the subject of sky surveys in the infrared \citep{sykes1992,reach2007,arendt2014} and recently also
in the visible wavelength range \citep{ishiguro2007}. 
Comet 103P/Hartley is the only comet identified in our analysis which is contained in any 
 of these surveys but unfortunately its trail remained undetected \citep{reach2007}.

We also performed IMEX simulations for the DESTINY$^+$ 
 (Demonstration and Experiment of Space Technology for INterplanetary voYage with Phaethon fLyby and dUst Science)
 mission to the asteroid (3200) Phaethon \citep{kawakatsu2013,arai2018} scheduled for  launch in 2024. Our simulations 
 show that along its presently planned interplanetary trajectory the  spacecraft will cross the trail of comet \honda, 
 giving the 
 Dust Analyzer on board \citep{kobayashi2018b} the possibility to detect and analyze 
 individual particles originating from this comet. 

For our IMEX simulations we used a candidate trajectory for  the interplanetary transfer of MMX to Mars
which is the current nominal spacecraft trajectory \citep{campagnola2018}. Any change in the MMX launch date will affect the 
actual  trajectory, and thus likely lead to a different set of cometary trail traverses. 
On the other hand, for the Mars orbital phase we used Mars as the target and these simulation results are  
valid independent of the spacecraft under consideration. They may  give  useful information about the occurrence of
meteors also for other  missions 
carrying landers or rovers to the red planet. 
For MMX they remain useful as long as  the spacecraft's actual operational phase 
at Mars is at least partially covered by our simulations. 

The fact that we performed our  simulations for the MMX Martian orbital phase  with Mars as a target
 may lead to 
 deviations in the impact directions and speeds as compared to those given in Table~\ref{tab:results} because
the spacecraft will
 be in an orbit very similar to the one of Phobos. 
Considering Phobos'  orbital 
 speed of about $\mathrm{2\, km\,s^{-1}}$, the uncertainty  in the impact speed caused by this approximation should be below 20\%. 
 Furthermore, considering that CMDM is a flat sensor with $2\pi$ field-of-view, the error in the impact direction should not lead to a 
 significant uncertainty in the dust flux, in particular when taking into account that the fluxes predicted by IMEX have
 an uncertainty of at least a factor of ten anyway. For a more detailed 
 analysis of the impact directions and speeds, and in particular for planning the CMDM measurements at Mars,
 it will be worthwhile to use the actual 
 MMX trajectory around Mars in the future.

Cometary meteoroid impacts may pose significant risks for Earth orbiting satellites and crewed 
spaceflight  as has been already mentioned in the Introduction. For crewed missions to other planets 
this will likely become an increasingly severe issue because the spacecraft will become increasingly larger
and  long flight times will be required which will increase the probability  for  hazardous meteoroid hits. \citet{christou2010b} 
pointed out that in the future the impact hazard should already be considered during the  analysis and design phase 
of interplanetary missions in order to optimize their trajectories
and reduce hazardous cometary trail traverses as much as possible.

\section{Conclusions}

 \label{sec:conclusions}

We used the IMEX dust streams in 
space model \citep[][]{soja2015a} to predict   cometary meteoroid trail traverses by the MMX spacecraft.  The model simulates  cometary  trails  
in the inner Solar System formed by particles in the size range $\mathrm{10^{-8}\,kg}$ to $\mathrm{10^{-2}\,kg}$, approximately corresponding to 
$\mathrm{100\, \mu m}$ to 1~cm particle radius. 

Using the presently available nominal trajectory of MMX we found that 
during its interplanetary transfer to Mars and while being in Mars orbit the  spacecraft 
will traverse 
several cometary meteoroid trails. The highest dust flux is 
predicted for October 2026 when MMX will be in Mars orbit traversing the trail of comet 114P/Wiseman-Skiff, with a maximum flux of 
$\mathrm{0.15\,m^{-2}\,day^{-1}}$ lasting for about three days. We find a total of 12 trail traverses during the entire mission, 
with fluxes of at least
$\mathrm{10^{-3}\,m^{-2}\,day^{-1}}$, covering
a total time period of approximately 90~days.  Smaller particles -- which are  not covered by the present IMEX model -- 
are expected to exist in the comet trails as well and may lead to a few orders of magnitude higher  dust fluxes. We conclude that the large area CMDM 
dust sensor on board MMX with its 
detection threshold of approximately $\mathrm{3\, \mu m}$ particle radius 
 will likely be able to detect trail particles during a few meteoroid trail traverses. 

 The in situ detection and analysis of cometary trail particles in space creates a new opportunity to remotely measure the composition 
 of celestial bodies without the necessity to fly a spacecraft to or even land on the source objects. Missions carrying dust instruments in the near future 
 that may benefit from this approach include  DESTINY$^+$  
 to the asteroid (3200) Phaethon, Europa Clipper to the Jupiter system  and the Interstellar Mapping and Acceleration Probe
 \citep{kobayashi2018b,kempf2018,mccomas2018}. Successful measurements of meteoroid trails by 
 spaceborne instruments may lead to improvements of cometary trail models. Forecasts of   trail traverses by  planets 
 in connection with in situ or remote spacecraft investigations can also
 provide new insights into atmospheric and ionospheric processes.  Last but not least, 
 they can make   essential predictions for spacecraft 
 hazards for future crewed deep space missions. 

\section*{Acknowledgements}

The IMEM model and the IMEX Dust Streams in Space model were
developed under ESA funding (contracts 21928/08/NL/AT and 4000106316/12/NL/AF -
IMEX). We are grateful to the MPI f\"ur Sonnensystemforschung, Chiba Institute of Technology and the University of Stuttgart for their support. H. Kimura 
gratefully acknowledges support by the Grants-in-Aid for Scientific Research (KAKENHI number 19H05085) of the Japan Society for the Promotion of Science (JSPS).
 We are grateful to two anonymous referees whose comments substantially improved the presentation of our results.

\clearpage 

\section{Appendix}

\begin{table}[htbp]
   \caption{Orbital elements of comets discussed in this paper from the JPL Small Body Database (ssd.jpl.nasa.gov). 
}
   \centering
   \small
   \begin{tabular}{@{} lcccccccc @{}} 
         \hline
\multicolumn{1}{c}{Comet}               & $e$  &   $q$  &    $i$     & $\Omega$   & $\omega$   &  $t_{\mathrm{Perihelion}}$ & Orbital Period & Epoch  \\  
                                        &      &  [AU]  &[$^{\circ}$]&[$^{\circ}$]&[$^{\circ}$]&                            & [yr]        &     \\
\multicolumn{1}{c}{(1)}                 &  (2) &   (3)  &    (4)     &     (5)    &     (6)    &       (7)    &     (8)     &     (9)   \\
   \hline
 3D/Biela                               & 0.75 & 0.88   &  13.2      &   250.7    &  221.7     & 26.11.1832   &      6.65   &     03.12.1832 \\
 19P/Borrelly                            & 0.62 & 1.36   &  30.3      &    75.4    &  353.4     & 14.09.2001   &      6.86   &     01.05.2004 \\
 21P/Giacobini-Zinner                   & 0.71 & 1.01   &  32.0      &   195.4    &  172.8     & 10.09.2018   &      6.55   &     05.11.2017\\
 24P/Schaumasse                         & 0.70 & 1.21   &    11.7    &    79.7    &  58.0      & 16.11.2017   &      8.26   &     24.11.2015 \\
 \hondaa                                & 0.82 & 0.53   &    4.2     &    89.0    &  326.3     & 31.12.2016   &      5.26   &     14.02.2017  \\
 103P/Hartley                           & 0.69 & 1.06   &  13.6      &   219.8    &  181.2     & 28.10.2010   &      6.46   &     30.11.2010 \\
 114P/Wiseman-Skiff                     & 0.55 & 1.58   &  18.3      &   271.1    &  172.8     & 14.01.2020   &      6.67   &     13.03.2018 \\ 
 185P/Petriew                           & 0.70 & 0.93   &  14.0      &   214.1    &  181.9     & 13.08.2012   &      5.46   &     30.06.2013 \\
 255P/Levy                              & 0.67 & 1.00   &  18.3      &   279.7    &  179.7     & 02.05.2017   &      5.30   &     20.11.2014 \\
  275P/Hermann                          & 0.71 & 1.65   &  21.6      &   349.3    &  173.1     & 19.02.1999   &     13.84   &     06.03.2004 \\
      \hline
   \end{tabular}
   \label{tab:comets}
\end{table}


\clearpage

\section*{Figure Captions}

\begin{figure}[htb]
\vspace{-2.2cm}
\hspace{-1.2cm}
		\includegraphics[width=1.16\textwidth]{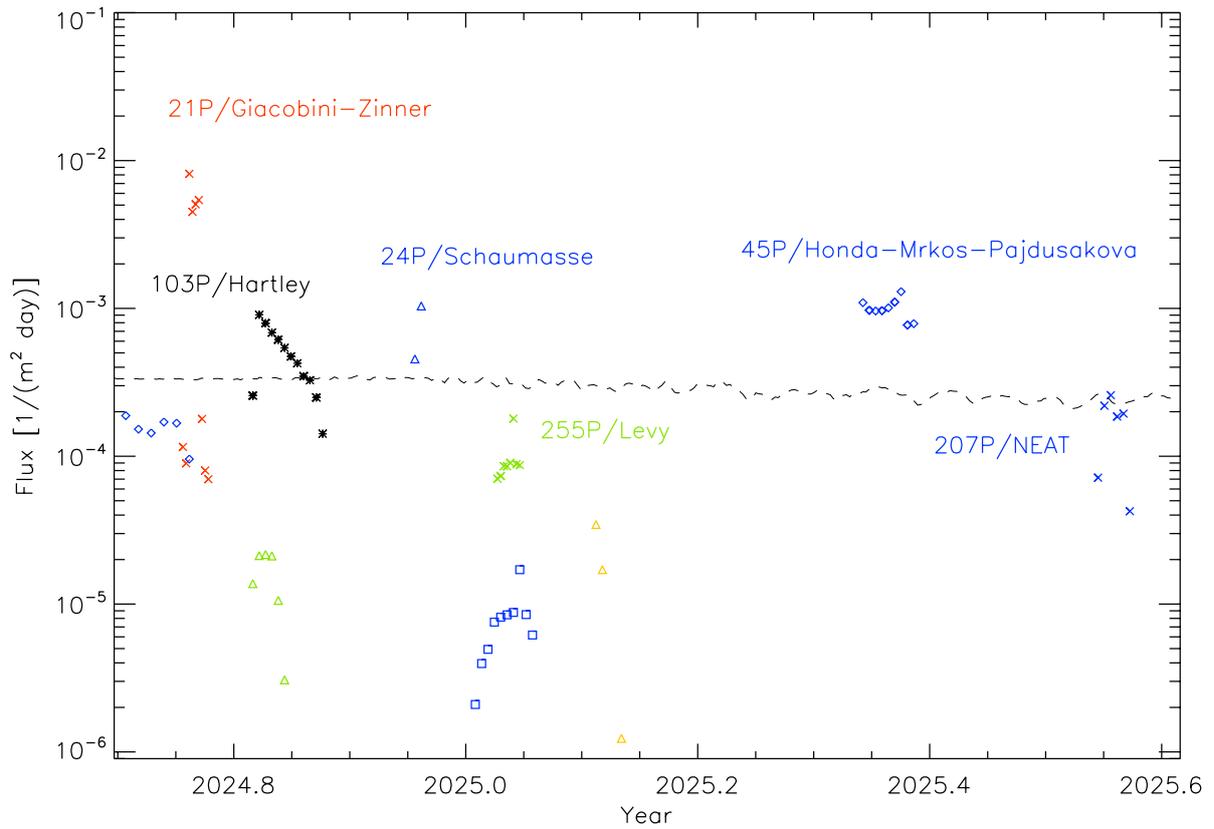}
		\vspace{-1.7cm}
	\caption{Simulated dust fluxes for cometary meteoroid trails intercepted by the MMX spacecraft during its interplanetary transfer
	to Mars (cf. Fig.~\ref{fig:orbitplot}). Simulated 
	particles are larger than $\mathrm{100\,\mu m}$. 
        Symbols and colours distinguish individual comets: 
        red crosses: 21P/Giacobini-Zinner;  
        blue triangles: 24P/Schaumasse; 
        blue diamonds: \honda; 
        black asterisks: 103P/Hartley; 
        green crosses: 255P/Levy; 
        blue crosses: 207P/NEAT; 
        green triangles: 249P/LINEAR; 
        blue squares: 263P/Gibbs; 
        yellow triangles: 41P/Tuttle-Giacobini-Kres{\'a}k. The simulations  
        were performed 
	with a 1-day step size for 21P, and with a 2-day  step size for all other comets. 
	The dashed line shows the simulated flux of sporadic interplanetary dust particles, see text for details.  
}
	\label{fig:flux}
\end{figure}

\begin{figure}[tbh]
\vspace{-0.7cm}
\hspace{-0.1cm}
	\vspace{-8.7cm}	
		\includegraphics[width=1.\textwidth]{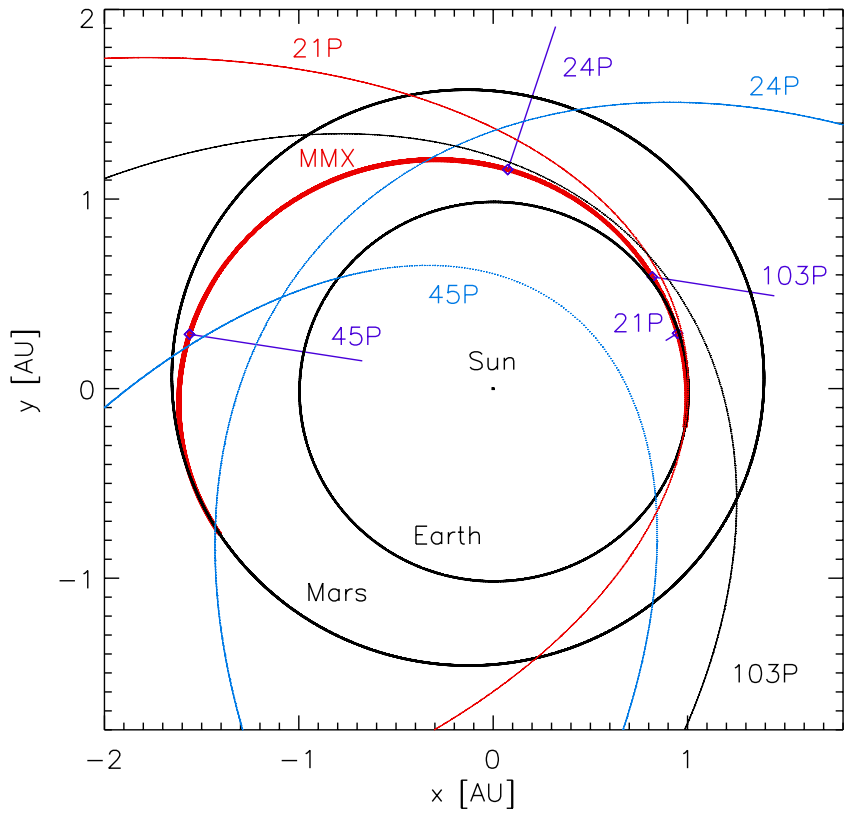}
	\caption{Interplanetary trajectory of MMX (thick red line). Blue diamonds indicate the crossings of  cometary  trails 
	with predicted dust fluxes of at least $\mathrm{10^{-3}\,m^{-2}\,day^{-1}}$ 
	(cf. Figure~\ref{fig:flux}).
	Blue lines indicate the approach direction (speed vector) of trail particles onto the spacecraft in the 
	spacecraft-centered reference frame  projected onto the X-Y plane
	(cf. Table~\ref{tab:results}). 
	The X-Y plane is the ecliptic plane with vernal equinox oriented towards the +X direction. The comet orbits 
	are superimposed as thin lines.}
	\label{fig:orbitplot}
\end{figure}

\begin{figure}[tbh]
\vspace{-2.cm}
\hspace{-1.4cm}
		\includegraphics[width=1.18\textwidth]{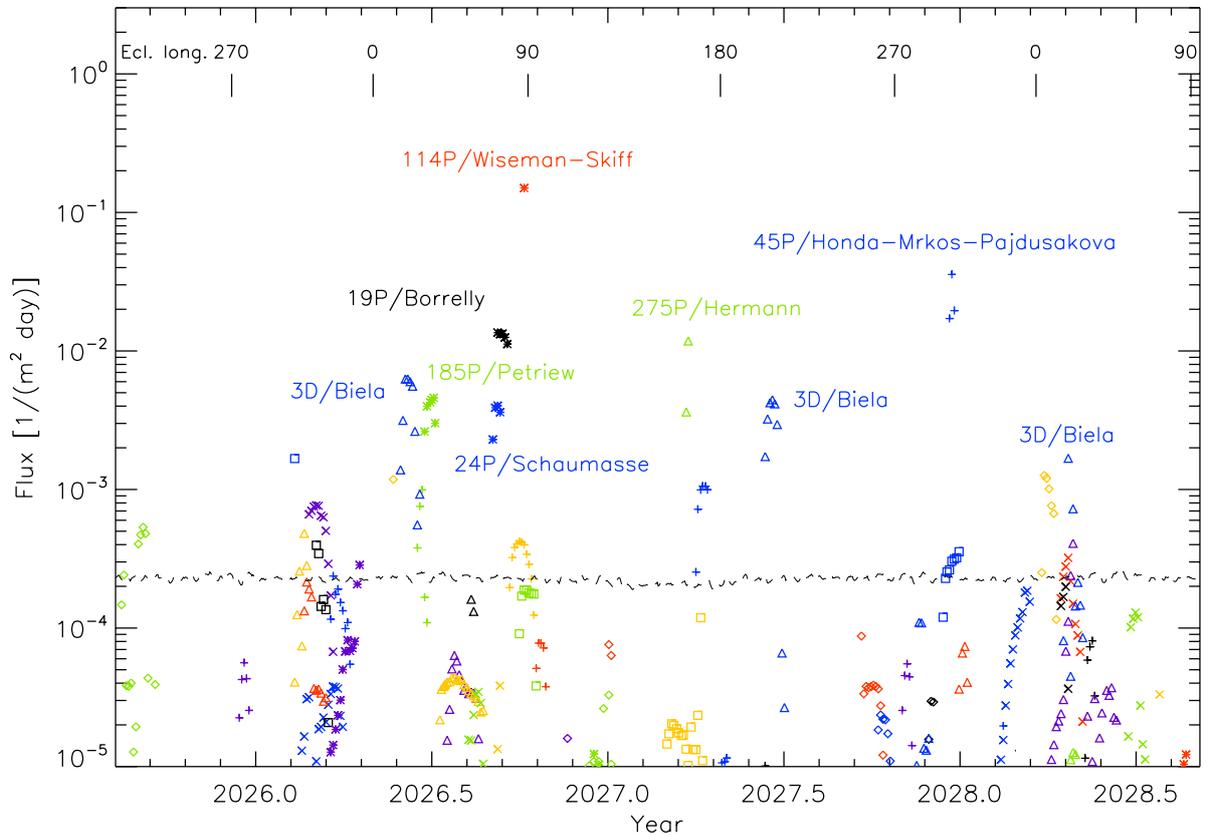}
		\vspace{-1.6cm}
				\caption{Simulated dust fluxes for cometary meteoroid trails intercepted by the MMX spacecraft during its orbital phase at
	 Mars (cf. Fig.~\ref{fig:trajectory_mars}). Simulated 
	particles are larger than $\mathrm{100\,\mu m}$. 
        Symbols and colors distinguish individual comets: red asterisks: 114P/Wiseman-Skiff;  black asterisks: 
        19P/Borrelly; blue triangles: 3D/Biela; 
        blue crosses: \honda; 
        green triangles: 275P/Hermann; green asterisks: 185P/Petriew; blue asterisks: 24P/Schaumasse; blue squares: 10P/Tempel~2. The remaining symbols 
        refer to other comets forming a 
        very low background flux. The simulations were performed 
	with a 2.5-day step size. The ecliptic longitude of Mars is indicated at the top, and the dashed line shows the simulated flux of 
	interplanetary dust particles, see text for details.  
}
	\label{fig:results}
\end{figure}

\begin{figure}[tbh]
\vspace{-0.6cm}
\hspace{-0.5cm}
	\vspace{-8.8cm}	
		\includegraphics[width=\textwidth]{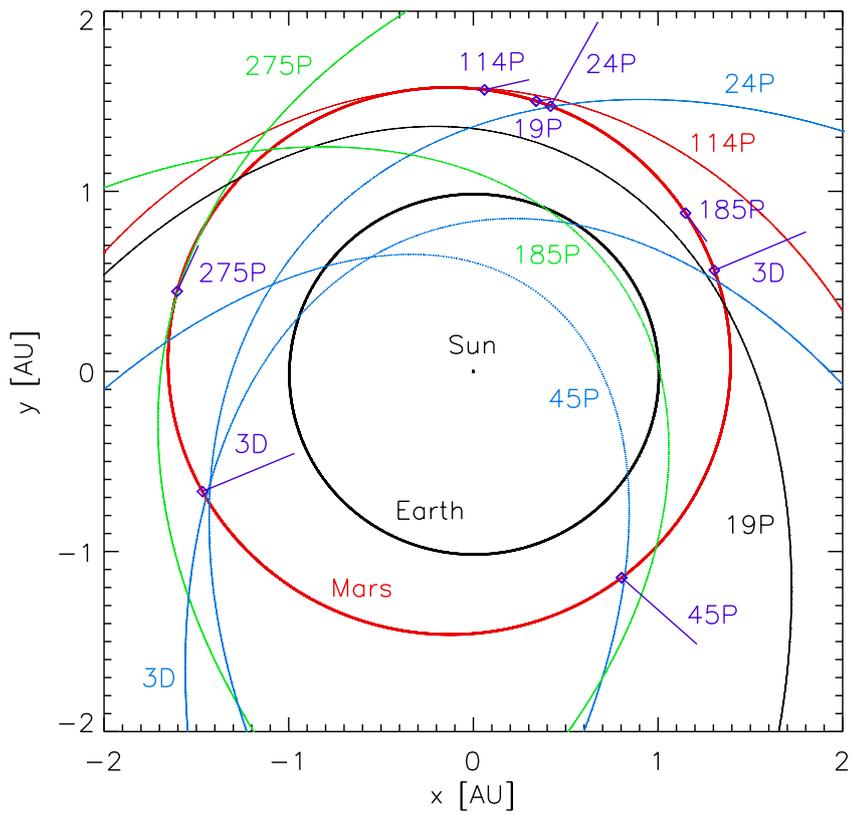}
	\caption{Same as Figure~\ref{fig:orbitplot} but for the Mars orbital phase of MMX.}
	\label{fig:trajectory_mars}
\end{figure}

\end{document}